\documentclass[
reprint,
superscriptaddress,
nofootinbib,
longbibliography,
bibnotes,
amsmath,amssymb,
aps,
prb,
]{revtex4-2}
\usepackage{graphicx}
\usepackage{bm}% bold math
\usepackage{dsfont}
\usepackage[usenames,dvipsnames]{xcolor}
\usepackage{pstricks}
\usepackage[tight]{subfigure}
\usepackage{verbatim}
\usepackage{units}
\usepackage{multirow}
\usepackage{enumitem}
\usepackage{mathrsfs}
\usepackage{leftidx}
\usepackage{xspace}
\usepackage{braket}
\usepackage{bbm}
\usepackage{cancel}
\usepackage{comment}
\usepackage{amsfonts,amssymb, amsmath}
\usepackage[normalem]{ulem}
\usepackage{hyperref}
\usepackage{appendix}
\hypersetup{colorlinks=true,linktoc=all,linkcolor=blue,breaklinks=true,citecolor=blue,urlcolor=blue}
\usepackage{orcidlink}
\usepackage[english]{babel}
\begin{document}
\title{Role of particle density in the Hall response of synthetic fermionic ladders: Lifshitz and Meissner-vortex transitions}

\author{Matteo Ferraretto~\orcidlink{0000-0003-0859-6681}}
\thanks{These authors contributed equally.}
\affiliation{Scuola Internazionale Superiore di Studi Avanzati, Via Bonomea 265, 34136, Trieste, Italy}
\author{Matteo Acciai~\orcidlink{0000-0001-9739-9289
}}
\thanks{These authors contributed equally.}
\affiliation{Scuola Internazionale Superiore di Studi Avanzati, Via Bonomea 265, 34136, Trieste, Italy}
\affiliation{The Abdus Salam International Centre for Theoretical Physics, Strada Costiera 11, 34151 Trieste, Italy}
\author{Massimo Capone~\orcidlink{0000-0002-9811-5089}}
\affiliation{Scuola Internazionale Superiore di Studi Avanzati, Via Bonomea 265, 34136, Trieste, Italy}
%\affiliation{CNR-IOM Democritos, Via Bonomea 265, I-34136 Trieste, Italy}
	
\date{\today}

\begin{abstract}
We characterize the Hall response of non-interacting fermionic $M$-leg ladders in the presence of an artificial magnetic flux, that can be realized in one-dimensional optical lattices supplemented with a synthetic dimension.

We focus on the Hall imbalance, which can be directly measured in experiments with ultracold fermionic atoms.
At relatively large synthetic flux we find a dependence of the density that contrasts with previously reported density-independent behavior. In particular, the Hall imbalance can be significantly enhanced in the limit of small density of particles (or holes), or it can vanish and change sign at specific fluxes, which we obtain analytically in the large inter-leg coupling regime.
This behavior is explained in terms of Lifshitz transitions of the band structure, where the number of Fermi points changes as a function of the flux and density. 
Finally, we explore the connection between these transitions and the so-called Meissner-vortex transition for fermionic ladders by computing the
site-resolved leg and rung currents and discussing
the similarities and differences with the bosonic counterpart.
\end{abstract}

\maketitle

\section{Introduction}
In the last three decades, platforms of ultracold gases trapped in optical lattices have proved to be a useful and flexible playground for the investigation of quantum systems in a controlled environment, where quantum simulators of model Hamiltonian with tunable parameters can be realized~\cite{Bloch_ManyBody_review, Block_quantum_simulation, Atomtronics_review}.
In particular, increasing attention has been devoted to the experimental realization of the Harper-Hofstadter model in a two-dimensional geometry \cite{Aidelsburger_2013, Chern_number_Hofstadter, Ketterle_Harper_Hofstadter} or in a ladder geometry \cite{Celi_synthetic_dimensions, chiral_currents_Science,Stuhl2015Sep,Tai2017Jun}, where a one dimensional system is supplemented with a synthetic short dimension provided by an internal atomic degree of freedom~\cite{Fallani_synthetic_dimension, Fallani_ladder_with_clock_transition} and an effective magnetic field can be induced either by Floquet engineering or by laser-assisted electronic transitions~\cite{Goldman_synthetic_gauge}. 

In the case of fermionic atoms, which is the main focus of the present work, a particularly promising experimental platform is provided by alkaline-earth-like atoms, such as $^{173}$Yb~\cite{Yb_SU(N)} and $^{87}$Sr~\cite{Sr_SU(N)}; these isotopes present a large number of nuclear states and a peculiar SU($N$) symmetric interaction~\cite{Gorshkov_SU(N), Cazalilla_2014}, which has recently been used to observe a flavor-selective Mott transition~\cite{Del_Re_selective_Mott, flavor_selective_experiment} and to probe the spin-exchange dynamics~\cite{Scazza_spin_exchange, Fallani_spin_exchange, Zhang_OFR}.
With this setup, the synthetic tunneling can be realized by laser-induced stimulated Raman electronic transitions to the state $^3P_1$ that, unlike the ground state $^1S_0$, displays hyperfine coupling, thus inducing an effective exchange of quanta of angular momentum between the nucleus and the photon field.
The phase gradient of the Raman laser beams along the optical lattice allows a particle to accumulate a phase when tunneling around a closed synthetic plaquette, mimicking the effect of a magnetic field applied perpendicularly to a ladder of electrically charged particles.
%In these ladder systems, the peculiar interaction schemes~\cite{Gorshkov_SU(N)}, particularly rich in the case of fermionic atoms, offer the possibility to investigate a plethora of phenomena, such as atoms a flavor-selective Mott transition~\cite{Del_Re_selective_Mott, flavor_selective_experiment}, spin-exchange dynamics \cite{Scazza_spin_exchange, Fallani_spin_exchange, Zhang_OFR}.
%For example, it is possible to realize synthetic (1+1)-dimensional ladders of effectively spinless fermions, where one ``real" dimension is provided by the optical lattice and the other ``synthetic" dimension is associated with the nuclear degree of freedom (flavor). 
The presence of an artificial gauge field acting on neutral atoms provides the system with intriguing properties, such as the presence of persistent chiral edge currents~\cite{chiral_currents_Science, Hugl_2014} that can be enhanced by interactions~\cite{ferraretto_scipost, Barbarino_2016} and, most importantly for the present work, offer the opportunity to simulate the Hall effect in a controlled environment.
%%%%%%%%%%%%%%%%%%%%%%%%%%%%%%%%%%%%%%%
\begin{figure}[t]
    \centering
    \includegraphics[width=0.80\linewidth]{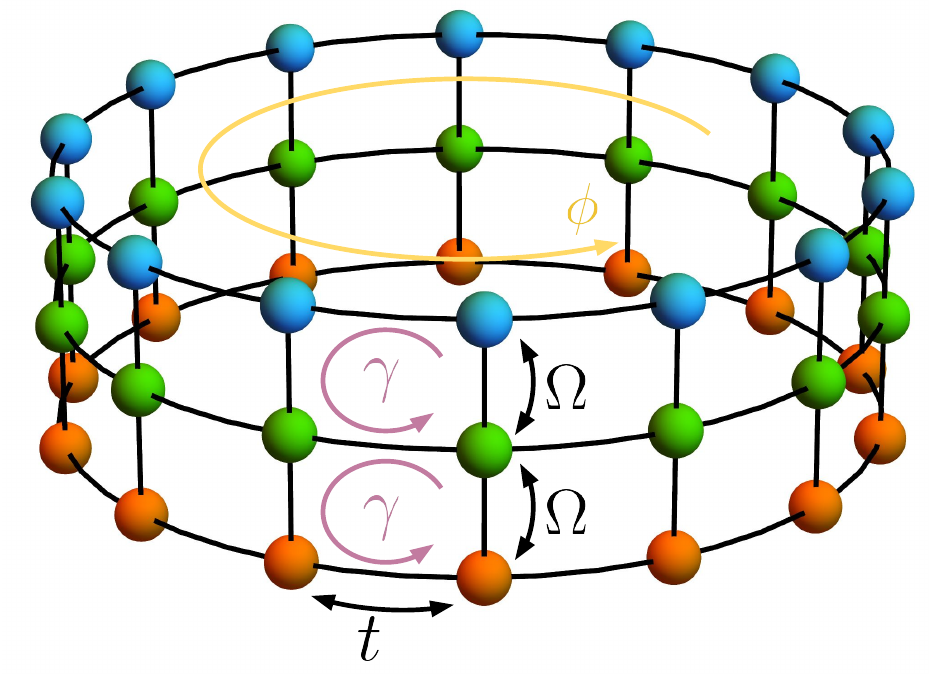}
    \caption{Sketch of a synthetic ladder with real PBC subject to an external flux $\phi$ piercing the ring and inducing a persistent current along the real direction.
    The synthetic direction is represented by different colors.
    }
    \label{fig:system}
\end{figure}
%%%%%%%%%%%%%%%%%%%%%%%%%%%%%%%%%%%%%%%

The Hall response in atomic synthetic ladders has been addressed theoretically \cite{Filippone_Giamarchi_Hall_imbalance, Giamarchi_prl_2021,Citro2025Feb} and probed in recent experiments \cite{Fallani_universal_Hall_response, Zhou2025Nov}.
One of the most striking results of these works is that the Hall imbalance, the ratio between polarization along the synthetic direction and total current along the real direction, computed and measured in the presence of interaction and a small finite temperature, displays a universal behavior, i.e., it is nearly independent of the particle density, interaction strength (above a certain threshold)~\cite{Filippone_Giamarchi_Hall_imbalance,Fallani_universal_Hall_response} and temperature, and it can be computed from the corresponding non-interacting system.
More specifically, the universal regime is achieved in the limit of large synthetic tunneling, where the bands are separated by sufficiently large gaps and in the limit of small gauge fluxes. Deviations from the universal behavior have also been reported \cite{Giamarchi_prl_2021} for intermediate values of the flux in the presence of interactions. 
However, despite these significant progresses, a comprehensive study of the non-interacting system spanning the full parameter regime and a thorough investigation of the breaking of universality are still lacking.

In the present work, we move one step further in this direction and, while focusing on the regime of gapped bands, we investigate the Hall response in a non-interacting $M$-leg synthetic fermionic ladder in the whole regime of fluxes and particle densities.
In particular we find that, already in the non-interacting system, a large value of the gauge field leads to a  strong dependence of the Hall imbalance  on the particle density, in contrast with the universal-response regime.

We characterize the nonuniversal behavior with a semi-analytic approach and we explain it in terms of Lifshitz phase transitions, where the number of Fermi points changes, driven by flux or density.
As a result, deep in the nonuniversal regime, we do not find just quantitative corrections to the universal prediction, but we also find qualitative changes. Depending on the details of the system, the so-called Hall imbalance can change sign, reflecting a transition from particle-like to hole-like behavior, and even diverge for certain values of the flux.

The behavior across the Lifshitz transitions reminds of some aspects of the so-called Meissner-vortex transition discussed in bosonic ladders~\cite{Orignac2001Sep,Petrescu2013Oct,Atala,Piraud2015Apr,DiDio2015Apr,DiDio2015Aug,Kolley2015Sep,Greschner2015Nov,Orignac_2016,Impertro2025Jun}, but we provide evidence that the similarity does not reflect a deep connection between the two phenomena.

The manuscript is organized as follows: in Sec.~\ref{sec:Model} we introduce the model and the main observables of interest; in Sec.~\ref{sec:2_legs} we explore the exact solution for the 2-leg ladder, exploring the full parameter regime; in Sec.~\ref{sec:Perturbation_theory} we introduce a perturbative formalism valid for a large synthetic tunneling in a generic $M$-leg ladder and we report on three particular examples that allow us to illustrate all the aspects of the general framework. 
Then in Sec.~\ref{sec:Bond_and_rung_currents} we explore the connection between the Hall response and the Meissner-vortex transition in the context of fermions; finally Sec. \ref{sec:Conclusions} is devoted to concluding remarks and outlook.

\section{Model}
\label{sec:Model}

We consider a fermionic ladder with $L$ sites along the real dimension, and $M$ legs, corresponding to the synthetic dimension generated by the internal degree of freedom (nuclear spin in the cold-atom implementations).
The Hall response of this system can be investigated within the framework of linear response theory by using periodic boundary conditions (PBC) along the real direction and coupling the ladder to two external fields: an Aharonov-Bohm flux $\phi$ piercing the real-space ring~\cite{Scalapino_White_Zhang} and a polarizing field $\nu$ along the synthetic dimension. 
Since the particles are electrically neutral, these quantities are not associated to the electromagnetic field, but they are here used as mathematical tools to extract the relevant response functions, as detailed below.
This system is sketched in Fig.~\ref{fig:system} and it is described by the following Hamiltonian:
\begin{align}\label{eq:Hamiltonian}
    H =& - t\sum_{j=1}^{L} \sum_{\sigma=\sigma_{\mathrm{min}}}^{\sigma_{\mathrm{max}}} \left[ e^{-i\phi} c^{\dagger}_{j\,\sigma} c_{j+1\,\sigma} + \mathrm{h.c.} \right] \nonumber \\ 
    & + \Omega \sum_{j=1}^L \sum_{\sigma=\sigma_{\mathrm{min}}}^{\sigma_{\mathrm{max}}-1} \left[ e^{i\gamma j} c^{\dagger}_{j\,\sigma} c_{j\,\sigma+1} + \mathrm{h.c.} \right] \nonumber \\
    & + 2\nu \sum_{j=1}^L \sum_{\sigma=\sigma_{\mathrm{min}}}^{\sigma_{\mathrm{max}}} \sigma n_{j\,\sigma}.
\end{align}
Here, $t$ and $\Omega$ represent the tunneling amplitudes along the real and synthetic dimensions respectively, $\gamma$ is the synthetic flux per plaquette, $c^{(\dag)}_{j\,\sigma}$ is a fermionic operator that destroys (creates) a particle at the site $j\sigma$, and $n_{j\,\sigma} = c^{\dag}_{j\,\sigma} c_{j\,\sigma}$. 
The synthetic-dimension index runs from $\sigma_{\mathrm{min}}=-\frac{M-1}{2}$ to $\sigma_{\mathrm{max}}=+\frac{M-1}{2}$ at integer steps.

With a unitary transformation\footnote{In order for this transformation to be consistent with PBC, one must impose the constraint $\sigma\gamma L=2\pi m$, for some integer $m$ and all $\sigma$. When $M$ is even, this implies that the allowed values of the magnetic flux are of the form $\gamma=4\pi m/L$, while for odd $M$ one has $\gamma=2\pi m/L$.
In the original gauge, the condition imposed by PBC would be $\gamma = 2\pi m/L$ for any ladder~\cite{Orignac_2_legs_ladder}.
} $c_{j\sigma} \to  e^{i\sigma \gamma j} c_{j\sigma}$, the Hamiltonian becomes explicitly translation invariant, so we can write $H = \sum_k \Psi^{\dag}_k H_k \Psi_k$, where $k=-\pi+\frac{2\pi n}{L}$ ($n=0,\dots,L-1$), and $\Psi^{\dag}_k = ( c^{\dag}_{k\,\sigma_{\mathrm{min}}}, \dots , c^{\dag}_{k\, \sigma_{\mathrm{max}}})$. The $M\times M$ matrix $H_k$ is tridiagonal, where the element at position $(\sigma-\sigma_{\mathrm{min}}+1)$ on the main diagonal reads $T_{\sigma\sigma} \equiv 2\sigma\nu -2t\cos{(k+\sigma \gamma - \phi)}$, and all elements on the superdiagonal and subdiagonal are given by $\Omega$.
For the following, it is useful to introduce the average current along the real direction ($x$) and the average polarization along the synthetic direction ($y$):
\begin{equation}
\label{eq:current_polarization_definition}
\begin{split}
    J_x(\phi,\nu) & = -\left\langle \partial_{\phi} H \right\rangle = -\partial_{\phi}E_0, \\
    P_y(\phi,\nu) & = \left\langle \partial_{\nu}H \right\rangle = \partial_{\nu} E_0,
\end{split}
\end{equation}
where we have used the Hellmann-Feynman theorem and introduced the internal energy $E_0(\phi,\nu) = \langle H \rangle$~\cite{Prelovsek1999Oct, Zotos2000Jul}.

The Hall effect is observed when, in response to an induced current along the real direction $J_x$, or equivalently to a small Aharonov-Bohm flux $\phi$, the system develops a polarization $P_y$ along the synthetic direction due to the presence of a finite gauge flux $\gamma$. 
We can thus characterize the Hall effect in the linear response regime with the \emph{Hall imbalance}~\cite{Filippone_Giamarchi_Hall_imbalance}, the ratio between polarization and current computed at $\nu = 0$ and $\phi \neq 0$ (where $\phi$ is a small perturbation):
\begin{equation}\label{eq:Delta_H_definition}
    \Delta_H = 2t \lim_{\phi \to 0} \frac{P_y(\phi,\nu=0)}{J_x(\phi,\nu=0)} = 2t \frac{ \partial^2_{\phi\nu} E_0 |_{\phi=\nu=0}}{ \partial^2_{\phi\phi} E_0 |_{\phi=\nu=0}},
\end{equation}
where we have rescaled the ratio by $2t$ to make $\Delta_H$ dimensionless.
The last equality is obtained by expanding quadratically the ground-state energy $E_0$ in the variables $(\phi,\nu)$ around the unperturbed point $(0,0)$.
Another quantity frequently used to characterize the Hall response is the \emph{Hall coefficient} $R_H$, which is discussed in Appendix~\ref{sec:Hall_coefficient}.
This quantity comes from the classical theory of the Hall effect and is closely related to $\Delta_H$, so in the following we mainly focus on the latter.

We will carefully map the behavior of  $\Delta_H(n,\gamma)$ as a function of the synthetic flux $\gamma$ and the particle density per leg $n = N/L$, where $N$ is the total number of particles and $0 \leq n \leq M$. 
Throughout this work we only focus on the case where the bands are always separated by a gap and they are never partially filled at the same time.
The results obtained in this regime can then be used to understand the more complicated situation of overlapping bands.
%Not only this is experimentally achievable by sufficiently increasing the power of the Raman beams, but also the results that we obtain here can be used to understand qualitatively the case of overlapping bands.
As a consequence of the particle-hole symmetry of Eq.~\eqref{eq:Hamiltonian}, the Hall response obeys the following symmetry properties: $\Delta_H(M-n,\gamma) = - \Delta_H(n,\gamma)$ and $\Delta_H(n, 2\pi-\gamma) = - \Delta_H(n, \gamma)$.
In particular, these properties imply that $\Delta_H(n,\pi) = 0$ and $\Delta_H(\frac{M}{2}, \gamma) = 0$~\cite{Filippone_2019}.
Considering also the $2\pi$-periodicity with respect to $\gamma$, we can limit our analysis to $0<n\leq\frac{M}{2}$ and $0\leq \gamma \leq \pi$.

\section{analytical solution for $M=2$}
\label{sec:2_legs}

In the particular case $M=2$, several properties of the model are known~\cite{Narozhny2005Apr,Orignac_2_legs_ladder,Carr2006May}. 
This will allow us to calculate the Hall response exactly in a closed analytical form.
Diagonalizing $H_k$, one gets the two bands
\begin{align}\label{eq:bands_2legs_exact}
    \varepsilon_{\pm}(k, \phi, \nu) =& -2t\cos{(k-\phi)}\cos{\frac{\gamma}{2}} \nonumber\\
    &\pm \sqrt{\left(2t\sin{(k-\phi)}\sin{\frac{\gamma}{2}} + \nu \right)^2 + \Omega^2}\,.
\end{align}
The unperturbed bands $\varepsilon_{\pm}(k,0,0) \equiv \varepsilon_{\pm}(k)$ are gapped if 
$\Omega > 2t\cos{\frac{\gamma}{2}}$.
Moreover, when $\gamma$ is smaller than 
\begin{equation}
    \gamma_c = 2\cos^{-1}\left[ \frac{\Omega}{4t}\left( \sqrt{1+\left(\frac{4t}{\Omega}\right)^2} - 1 \right) \right],
\end{equation}
the bottom band features a single minimum at $k=0$. Otherwise,
the band develops two symmetric minima at $k=\pm k_0$ (whose precise values are not so important for what follows), and the state $k=0$ is a local maximum. 

Since as $\gamma \to \pi$ the energy of $k=0$ approaches the energy of $k=\pm \pi$, we conclude that for any given particle density $n$, there must be a value $\gamma^\star$ at which the number of Fermi points changes from 2 (if $\gamma < \gamma^\star$) to 4 (if $\gamma > \gamma^\star$).
This is a one-dimensional analogue of the topological Lifshitz transitions of the Fermi surface~\cite{Lifshitz_original_paper, Lifshitz_transition, Huang2022Mar}.
If $\gamma < \gamma^\star$ the 2 Fermi points are $k_F=\pm \alpha \pi$, where $\alpha$ is such that summing over occupied states gives the total number of particles, $\sum_{k\,\text{occ.}} 1 = N$.
In the thermodynamic limit, we can replace the sum with an integral and readily obtain $\alpha = n$.
On the other hand, if $\gamma > \gamma^\star$ the 4 Fermi points are $k_{F,1}=\pm \beta \pi$ and $k_{F,2}= \pm \alpha \pi$, where conventionally $\beta < \alpha$, and they obey the equations $\sum_{k\,\text{occ.}} 1 = N \implies \alpha-\beta = n$ and $\varepsilon_-(\alpha\pi) = \varepsilon_-(\beta\pi)$.
The value of $\gamma^\star$ is given by $\beta = 0$,
yielding
\begin{equation}\label{eq:gamma_star_2legs_exact}
    \gamma^\star = 2\cos^{-1}\left[ \frac{\Omega}{4t}\left( \sqrt{1+\left(\frac{4t}{\Omega} \cos{\frac{n\pi}{2}} \right)^2} - 1 \right) \right].
\end{equation}
We observe that (i) in the limit $n\to 0$, consistently $\gamma^\star \to \gamma_c$; (ii) in the opposite limit $n\to 1$, $\gamma^\star \to \pi$ and (iii) if $\Omega\gg t$, then $\gamma^\star \approx \pi - \frac{4t}{\Omega} \cos^2{\frac{n\pi}{2}} + \mathcal{O}(\frac{t^3}{\Omega^3} )$.

From Eq.~\eqref{eq:Delta_H_definition} we can now compute the Hall imbalance by evaluating the relevant derivatives of the internal energy.
For example, $\partial^2_{\phi\nu} E_0 |_{\phi=\nu=0} = \sum_{k\,\text{occ.}} \partial^2_{\phi\nu} \varepsilon_- |_{\phi=\nu=0}$, where the sum runs over the occupied states in the unperturbed system.
The result is
\begin{subequations}
\begin{align}
\Delta_H&=\Delta_H^0{[\mathcal{F}_1(n)-\mathcal{F}_2(n)]}^{-1},\,\,\gamma\le\gamma^\star,
\label{eq:Delta_H_two_legs_exact_a}\\
\Delta_H&=\Delta_H^0\frac{\sin(\alpha\pi)/\mathcal{F}_1(\alpha)-(\alpha\to\beta)}{\sin(\alpha\pi)[1-\frac{\mathcal{F}_2(\alpha)}{\mathcal{F}_1(\alpha)}]-(\alpha\to\beta)},\,\,\gamma>\gamma^\star.
\label{eq:Delta_H_two_legs_exact_b}
\end{align}
\end{subequations}

\begin{comment}
\begin{widetext}
\begin{subequations}
\begin{align}
    \label{eq:Delta_H_two_legs_exact_a}
    \Delta_H = \frac{\frac{2t}{\Omega}\tan\left(\frac{\gamma}{2}\right)}{\displaystyle{\sqrt{1 + \frac{4t^2}{\Omega^2}\sin^2(\pi n)\sin^2\left(\frac{\gamma}{2}\right)} -\frac{2 t}{\Omega} \cos (\pi n) \sin \left(\frac{\gamma }{2}\right)\tan\left(\frac{\gamma}{2}\right)}} 
    \;\;\;\;\; \gamma \leq \gamma^\star \\
    \label{eq:Delta_H_two_legs_exact_b}
    \Delta_H = \frac{2t}{\Omega}\tan\left(\frac{\gamma}{2}\right)\frac{ \frac{\sin{(\alpha\pi )} }{\sqrt{1 + \frac{4t^2}{\Omega^2}\sin^2{(\alpha \pi)}\sin^2{\frac{\gamma}{2}}}} - (\alpha \rightarrow \beta)}{\sin{(\alpha\pi)} \left(1 - \frac{2t\cos{(\alpha\pi)}\tan{\frac{\gamma}{2}} \sin{\frac{\gamma}{2}}}{\sqrt{\Omega^2 + 4t^2 \sin^2{(\alpha\pi)} \sin^2{\frac{\gamma}{2}} }} \right) - (\alpha \rightarrow \beta)} \;\;\;\;\; \gamma > \gamma^\star,
\end{align}
\end{subequations}
\end{widetext}
\end{comment}
\noindent
Here $(\alpha \rightarrow \beta)$ indicates the same expression at the left with $\alpha$ replaced by $\beta$, $\Delta_H^0=-\frac{2t}{\Omega}\tan\left(\frac{\gamma}{2}\right)$,
and we have defined the functions
\begin{align}
    \mathcal{F}_1(x)&=\sqrt{1 + \frac{4t^2}{\Omega^2}\sin^2(\pi x)\sin^2\left(\frac{\gamma}{2}\right)},\\
    \mathcal{F}_2(x)&=\frac{2 t}{\Omega} \cos (\pi x) \sin \left(\frac{\gamma }{2}\right)\tan\left(\frac{\gamma}{2}\right).
\end{align}
\begin{figure}
    \centering
    \includegraphics[width=1.0\linewidth]{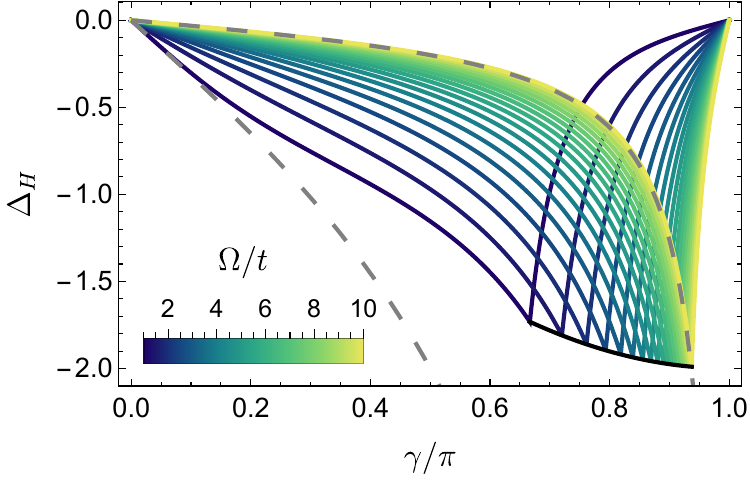}
    \caption{Hall imbalance as a function of the synthetic flux $\gamma$ in the 2-legs ladder with $n=1/2$ for several values of the synthetic tunneling $\Omega$. 
    The dashed gray lines represent the universal prediction $-\frac{2t}{\Omega}\tan{\frac{\gamma}{2}}$ for the two extreme values $\Omega=t$ and $\Omega=10t$.
    The solid black line marks the values of $\gamma^*$ and corresponding Hall imbalance obtained by varying $\Omega$. }
    \label{fig:2-legs-exact-vary-Omega}
\end{figure}
The result is shown in Fig.~\ref{fig:2-legs-exact-vary-Omega} for the representative case $n=1/2$.
First of all, in the limit $\Omega \gg t$ and $\gamma \ll \gamma^\star$ we recover the universal behavior $\Delta_H \approx -\frac{2t}{\Omega}\tan{\frac{\gamma}{2}}$ obtained in Ref.~\cite{Filippone_Giamarchi_Hall_imbalance} in the limit $\gamma \to 0$ and observed in Ref.~\cite{Fallani_universal_Hall_response} for intermediate values of the flux.
However, the exact results depart from this prediction as $\gamma$ approaches $\gamma^\star$ from below; the reason is that
in this regime $\tan{\frac{\gamma}{2}}$ becomes large and the second term at the denominator of Eq.~\eqref{eq:Delta_H_two_legs_exact_a} cannot be neglected, unless $\Omega/t$ is very small.
This introduces a dependence on the density and a breaking of the universal behavior at large fluxes.
Secondly, plugging Eq.~\eqref{eq:gamma_star_2legs_exact} into \eqref{eq:Delta_H_two_legs_exact_a}, we get the value of $\Delta_H$ at the critical point $\gamma^\star$. 
It is particularly interesting to expand this result in the limit $n\ll 1$: 
\begin{equation}
    \Delta_H=-\frac{\sqrt{2}}{\pi^2 n^2}\sqrt{\frac{\Omega}{t}  \left(\sqrt{\frac{\Omega ^2}{t^2}+16}-\frac{\Omega }{t}\right)}+\mathcal{O}(n^0),
\label{eq:deltaH_exact_2leg_small_n}
\end{equation}
which shows how $|\Delta_H|$ increases as the density decreases and eventually diverges as $n\to 0$.
Finally, as $\gamma \to \pi$, we can explicitly solve the equations for the Fermi points and obtain $\alpha = \frac{1+n}{2}$ and $\beta = \frac{1-n}{2}$; substituting these values into Eq.~\eqref{eq:Delta_H_two_legs_exact_b} we get $\Delta_H \to 0$ as expected.
In the interval $\gamma^\star < \gamma < \pi$ there is no analytic expression for the Fermi points\footnote{It is possible to obtain a closed expression at fixed chemical potential, but not at fixed density.}, and consequently we cannot write an explicit expression for $\Delta_H(n,\gamma)$.
However, numerical results shown in Fig.~\ref{fig:2-legs-exact-vary-Omega} show that  $\Delta_H$ increases monotonically in this interval; we characterize quantitatively this behavior in the limit $\Omega \gg t$ in Sec.~\ref{subsec:2-legs_ladder}.

\section{Perturbative approach to the $M$-leg ladder}
\subsection{General expressions}
\label{sec:Perturbation_theory}

For ladders with $M>2$ we cannot derive closed expressions for the exact eigenvalues of $H_k$. In this section we derive analytical expressions valid for $\Omega \gg t$ (treating $t$ as a perturbation) that will help us to interpret the numerical results in a wide range of values of $\Omega$.

The unperturbed bands at equilibrium, obtained by diagonalizing $H_k$ with $t=0$ and $\phi=\nu=0$, and the relative eigenvectors, whose components can be labeled by the leg index $\sigma$, are 
\begin{align}
    \varepsilon^{(0)}_p =& \, -2\Omega \cos{\left(\frac{(p+1)\pi}{M+1}\right)} \nonumber \\
    |p\rangle_{\sigma} =& \sqrt{\frac{2}{M+1}} \sin{\left( \frac{\pi}{2} \frac{(M-p)(M+1+2\sigma)}{M+1} \right)},
\end{align}
where $p=0,\dots, M-1$ labels the bands from the lowest to the highest.
Considering small external fields and treating the diagonal part of $H_k$, which has a term $\propto t$ and another $\propto \nu$, as the perturbation, we can derive the bands at second order:
\begin{align}\label{eq:bands_perturbation_theory}
    &\varepsilon^{(2)}_p(k,\phi,\nu) = \varepsilon^{(0)}_p + \sum_{\sigma} C_{pp\,\sigma} T_{\sigma\sigma}(k,\phi,\nu) \nonumber \\
    &+ \sum_{p^\prime \neq p} \frac{1}{\varepsilon^{(0)}_p - \varepsilon^{(0)}_{p^\prime}} \sum_{\sigma\rho} C_{pp^\prime\,\sigma} C_{pp^\prime\, \rho} T_{\sigma\sigma}(k,\phi,\nu) T_{\rho\rho}(k,\phi,\nu),
\end{align}
where $T_{\sigma\sigma}$ is defined before Eq.~\eqref{eq:current_polarization_definition}, and we introduced the coefficients $C_{pp^\prime\,\sigma} \equiv \langle p |_{\sigma} |p^\prime \rangle_{\sigma}$.
They satisfy the following relations: (i) normalization $\sum_{\sigma}C_{pp\,\sigma} = 1$; (ii) leg-reflection symmetry $C_{pp^\prime\,-\sigma} = (-1)^{p+p^\prime} C_{pp^\prime\,\sigma} $ and (iii) band-reflection symmetry $C_{M-1-p,M-1-p^\prime \, \sigma} = C_{pp^\prime\, \sigma}$.
From these relations, we can deduce important symmetry properties of the bands at equilibrium $\varepsilon^{(2)}_p(k,0,0)$.

For odd $M$ the third term in~\eqref{eq:bands_perturbation_theory} vanishes identically for the central band $p = \frac{M-1}{2}$. 
As a consequence, $\partial^2_{\phi\nu} E_0 = 0$ and the Hall imbalance vanishes identically for $\frac{M-1}{2} < n < \frac{M+1}{2}$. 
In this case, second-order perturbation theory is in general insufficient, and the Hall imbalance should be computed by keeping more terms in the expansion of the bands.
Remarkably, at some (band dependent) fluxes, the second term in \eqref{eq:bands_perturbation_theory} vanishes identically, and consequently the band is symmetric under $k\to \pi-k$.
This directly implies a vanishing polarization and consequently $\Delta_H = 0$.
These \emph{magic fluxes} are the solution of the trigonometric equation
\begin{equation}\label{eq:magic_equation}
    \sum_{\sigma} C_{pp\, \sigma} \cos{(\sigma\gamma)} = 0,
\end{equation}
namely 
\begin{equation}
\gamma_0=2\pi\frac{q}{M+1},\quad\!\! q=
    \begin{cases}
    1,2,\dots,\frac{M+1}{2} & M\text{ odd},\\
    1,2,\dots,\frac{M}{2},\frac{M+1}{2} & M\text{ even}.
    \end{cases}
\label{eq:magic_fluxes}
\end{equation}
In both cases, $1+p$ and $M-p$ are excluded from the allowed values of $q$.
In particular, the flux $\gamma_0 = \pi$ is always a solution for every $M$. Numerical calculations show that, except for $\gamma_0=\pi$ the symmetry under $k\to\pi-k$ does not hold for the exact band $\varepsilon_p(k)$, but only for the second-order expression $\varepsilon_p^{(2)}(k)$. Hence, the positions of the true zeros of $\Delta_H$ deviate  from those predicted by~\eqref{eq:magic_fluxes} as $\Omega$ decreases.
%As a corollary to these two properties, we observe that the central band in ladders with an odd number of legs is a flat band when computed at the magic fluxes within second order perturbation theory.

Within perturbation theory we can also compute analytically the Lifshitz points $\gamma^\star$ for $M>2$. The band $\varepsilon^{(2)}_p(k)$ has only either 2 or 4 Fermi points, unless it is a flat band (while the exact $\varepsilon_p(k)$ can have more than 4 Fermi points in some particular cases).
For small values of $\gamma$ there are 2 Fermi points for all the values of $M$ and $p$: we label these points with $k=\pm\alpha \pi$. 
Increasing $\gamma$ a Lifshitz transition ($L_1$) occurs and the band develops two extra Fermi points, which we label $k=\pm \beta\pi$.
Further increasing $\gamma$, more Lifshitz transitions can occur ($L_2$, $L_3$, etc.) and the band loses or gains two Fermi points at every phase transition; finally when $\gamma=\pi$ the band is symmetric under $k\to \pi-k$ and there are necessarily 4 Fermi points.
At a Lifshitz transition, the new points originate in $k=0$ or in $k=\pm \pi$, which means that we can identify the transition with one of the following 4 conditions: $\alpha = 0$, $\alpha = 1$, $\beta = 0$ or $\beta = 1$.
An example of evolution of the band and of the Fermi points induced by the flux is shown in Fig.~\ref{fig:4-legs-bands} for the bands $p=0$ and $p=1$ of the 4-leg ladder.
%Conventionally, in the phases with 2 Fermi points we set one of the following: $\alpha = 0$, $\alpha = 1$, $\beta=0$ or $\beta = 1$ in such a way that the functions $\alpha(\gamma)$ and $\beta(\gamma)$ are continuous.
%Neglecting the case of flat bands, at the critical points there are two possible situations: either the two extra Fermi points develop at $k=0$ ($\beta = 0$ as for the lowest band of the two-legs ladder), or they develop at $k=\pm \pi$ ($\beta = 1$).
%In the phase with 4 Fermi points, the occupied momentum states are the intervals $[-\alpha\pi, -\beta\pi] \cup [\beta\pi, \alpha\pi]$ in the former case and $[-\pi,-\beta\pi]\cup [-\alpha\pi,\alpha\pi]\cup [\beta\pi, \pi]$ in the latter.

The Fermi points can be obtained from the equations $\sum_{k\,\text{occ.}} 1 = N - pL$ and $\varepsilon^{(2)}_p(\alpha\pi) = \varepsilon^{(2)}_p(\beta\pi)$, where the latter is necessary only in presence of 4 Fermi points.
The Lifshitz transition flux can be obtained by imposing one of the 4 previously mentioned conditions,
i.e., by solving one of the following equations: 
\begin{equation}\label{eq:critical_equation}
    f_p(\gamma) = \left\{
    \begin{array}{lll}
        \frac{t}{\Omega} \sin^2{\left(\frac{n-p}{2}\pi\right)} & \alpha = 0, & \beta = 1-(n-p) \\
        -\frac{t}{\Omega} \cos^2{\left(\frac{n-p}{2}\pi\right)} & \alpha = 1, & \beta = 1-(n-p) \\
        \frac{t}{\Omega} \cos^2{\left(\frac{n-p}{2}\pi\right)} & \beta = 0, & \alpha = n-p \\
        -\frac{t}{\Omega} \sin^2{\left(\frac{n-p}{2}\pi\right)} & \beta = 1, & \alpha = n-p 
    \end{array}
    \right.
\end{equation}
%The first equation is $\sum_{k\,\text{occ.}} 1 = N - pL$, which in the phase with 2 Fermi points directly yields the solution $\alpha = n-p$ or $\beta = 1-(n-p)$, while in the phase with 4 Fermi points can be simplified as $\alpha-\beta = n-p$ if $\alpha>\beta$ or $\alpha-\beta = (n-p) - 1$ if $\beta > \alpha$.
%In the latter case, we also need the second defining equation $\varepsilon^{(2)}_p(\alpha\pi) = \varepsilon^{(2)}_p(\beta\pi)$, which can be recast as
%\begin{equation}\label{eq:defining_equation_Fermi_points}
   % \cos{\left(\frac{\alpha+\beta}{2}\pi\right)} = \frac{f_{p}(\gamma)}{\frac{t}{\Omega}\cos{\left(\frac{\alpha-\beta}{2}\pi\right)}},
%\end{equation}
where $f_{p}(\gamma)$ is a characteristic function of the given band:
\begin{equation}\label{eq:f}
    f_{p}(\gamma) \equiv \frac{\sum_{\sigma} C_{pp\sigma} \cos{(\sigma\gamma)}}{\sum_{p^\prime\neq p} \frac{4\Omega}{\varepsilon_p^{(0)} - \varepsilon_{p^\prime}^{(0)}} \sum_{\sigma\rho} C_{pp^\prime\sigma}C_{pp^\prime\rho} \cos{[(\sigma+\rho)\gamma]} }.
\end{equation}
Interestingly, this function vanishes at the magic fluxes: $f_p(\gamma_0) = 0$.
%The right-hand side of eq. \ref{eq:defining_equation_Fermi_points} only depends on $\alpha-\beta$, which can be replaced with either $n-p$ or $n-p-1$ to solve the equation and obtain an expression for $\alpha+\beta$, and consequently for $\alpha$ and $\beta$ separately.
%The Lifshitz critical fluxes $\gamma^\star$ can be computed by setting $\alpha = 0$, $\alpha=1$, $\beta=0$ or $\beta = 1$ in the defining equations of the Fermi points, which leads to the following equations:
Since $\Omega \gg t$, the right-hand side of Eq.~\eqref{eq:critical_equation} is small, hence $f_p(\gamma^\star)$ is small, which means that every Lifshitz flux $\gamma^\star$ is numerically close to one of the magic fluxes $\gamma_0$.

A general expression for the Hall imbalance can be obtained by computing the relevant second derivatives of the internal energy. For example if $p<n<p+1$ (i.e., all the bands with $p^\prime = 0,\dots,p-1$ are completely filled and the $p$-th band is partially filled) then only the $p$-th band contributes to the second derivatives
%: $\partial^2_{\nu\phi} E_0 |_{\phi=\nu=0} = \sum_{k\,\text{occ.}} \partial^2_{\nu\phi} \varepsilon^{(2)}_p|_{\phi=\nu=0} $ and $\partial^2_{\phi\phi} E_0 |_{\phi=\nu=0} = \sum_{k\,\text{occ.}} \partial^2_{\phi\phi} \varepsilon^{(2)}_p|_{\phi=\nu=0} $ and the result is
and the result is
\begin{widetext}
\begin{equation}\label{eq:Delta_H_perturbation_theory}
    \Delta_H \approx -\frac{\sum_{p^\prime\neq p} \frac{8t}{\varepsilon^{(0)}_{p^\prime} - \varepsilon^{(0)}_p} \sum_{\sigma}  \sigma C_{pp^\prime \sigma} \sum_{\rho} C_{pp^\prime \rho}\sin{(\rho\gamma)} \sum_{k\;\text{occ.}} \cos{k}}{\sum_{\sigma}C_{pp\sigma}\cos{(\sigma\gamma)} \sum_{k\;\text{occ.}} \cos{k} + \sum_{p^\prime \neq p} \frac{4t}{\varepsilon^{(0)}_{p^\prime} - \varepsilon^{(0)}_p} \sum_{\sigma\rho} C_{pp^\prime \sigma} C_{pp^\prime \rho} \cos{[(\sigma+\rho)\gamma]} \sum_{k\;\text{occ.}} \cos{(2k)} }\,.
\end{equation}
\end{widetext}
The density dependence of this expression is implicit in the sums over $k$, which can be computed explicitly in the continuum limit.
Let $I_1 = \frac{2\pi}{L} \sum_{k\,\text{occ.}} \cos{k}$ and $I_2 = \frac{2\pi}{L} \sum_{k\,\text{occ.}} \cos{2k}$, then with 2 Fermi points we have
\begin{align}\label{eq:integrals_2_Fermi_points}
    I_1 = 2\sin{(n\pi)},
    \;\;\;\;\;\;\;\;
    I_2 = \sin{(2n\pi)};
\end{align}
while with 4 Fermi points we have
\begin{align}\label{eq:integrals_4_Fermi_points}
    I_1 =& \frac{2\Omega}{t} \sin{[(\alpha-\beta)\pi]} f_p(\gamma), \nonumber\\
    I_2 =& 2\sin{[(\alpha-\beta)\pi]} \left[\frac{\frac{2\Omega^2}{t^2} f^2_p(\gamma)}{ \cos^2{\left(\frac{\alpha-\beta}{2}\pi\right)}} - 1 \right], 
\end{align}
where $\alpha-\beta$ has to be replaced by $n-p$ or $n-p-1$ depending on the shape of the band.

If $\gamma$ is far from any of the magic fluxes, the first term at the denominator of Eq.~\eqref{eq:Delta_H_perturbation_theory} is non-vanishing and dominant on the second term, which is of order $\mathcal{O}(\frac{t}{\Omega})$ and can be neglected.
Furthermore, we can simplify the factor $I_1$ and the final result is thus independent of the density; in other words the Hall imbalance shows the universal behavior.
On the other hand, if $\gamma \approx \gamma_0$ the first term at the denominator is small and it can be expanded in powers of $\frac{t}{\Omega}$, so in principle we cannot neglect the second term and $\Delta_H$ is therefore strongly dependent on the density, thus breaking the universal behavior.
Although it is not evident from Eq.~\eqref{eq:Delta_H_perturbation_theory} $\Delta_H = 0$ exactly at the magic fluxes $\gamma_0$ because the numerator vanishes.

In the following, we apply the general result~\eqref{eq:Delta_H_perturbation_theory} to the cases $M=2,3,4$, experimentally accessible values~\cite{Celi_synthetic_dimensions, chiral_currents_Science,Zhou2025Nov} which 
help us illustrate the main features of the perturbative theory and to highlight the different properties of $\Delta_H$ as $M$ is varied.
%Furthermore, the experimental realization of the Hall effect in such small ladders is within the reach of current experimental platforms~\cite{Celi_synthetic_dimensions, chiral_currents_Science,Fallani_arxiv}.
%In the following we apply this formalism to three particular cases $M=2,3,4$, which show a variety of different situations, allowing us to show practical examples of all the features of this model.

\subsection{2-leg ladder}
\label{subsec:2-legs_ladder}

To begin with, let us consider a 2-leg ladder with $0<n<1$, i.e., where only the lowest band labeled by $p=0$ is partially filled.
The only magic flux for this band is $\gamma_0 = \pi$ and the characteristic function is $f_0(\gamma) = \cos{\frac{\gamma}{2}} / [2 \sin^2{\frac{\gamma}{2}}]$. 
Since at the Lifshitz transition the extra Fermi points originate at $\beta = 0$, Eq.~\eqref{eq:critical_equation} reduces to $f_0(\gamma) = \frac{t}{\Omega} \cos^2{\left(\frac{n\pi}{2}\right)}$. 
%; we can see this from the general formula $\gamma_0 = 2\pi \frac{x}{M+1} = 2\pi \frac{x}{3}$, where since $M$ is even, $x$ takes the values $x=1$ or $x=\frac{3}{2}$ that are different from $1+p=1$ or $M-p=2$, i.e. only $x=\frac{3}{2}$.
%From the coefficients $C_{00\, \pm \frac{1}{2}} = \frac{1}{2}$ we get that the characteristic function of this band is $f_0(\gamma) = \cos{\frac{\gamma}{2}} / [2 \sin^2{\frac{\gamma}{2}}]$ and since the extra Fermi points develop at $\beta = 0$, the equation for the Lifshitz flux $\gamma^\star$ is $f_0(\gamma) = \frac{t}{\Omega} \cos^2{\left(\frac{n\pi}{2}\right)}$. 
However, since in general $\gamma^\star \approx \gamma_0 = \pi$, we can expand the characteristic function around the magic flux: $f_0(\gamma) \approx \frac{\pi-\gamma}{4} + \mathcal{O}(\pi-\gamma)^3$ and we obtain $\gamma^\star \approx \pi - \frac{4t}{\Omega} \cos^2{\left(\frac{n\pi}{2} \right)}$, which is consistent with the large-$\Omega$ expansion of Eq.~\eqref{eq:gamma_star_2legs_exact}.

We can now compute $\Delta_H$ from Eq.~\eqref{eq:Delta_H_perturbation_theory}. 
If $\gamma < \gamma^\star$ there are 2 Fermi points, the sums over $k$ are given in Eq.~\eqref{eq:integrals_2_Fermi_points} and the result is:
\begin{equation}\label{eq:Delta_H_perturbative_2_legs_2_Fermi_points}
    \Delta_H =-\frac{\frac{2t}{\Omega} \tan{\frac{\gamma}{2}}}{1- \frac{2t}{\Omega} \sin{\frac{\gamma}{2}}\tan{\frac{\gamma}{2}}\cos{(n\pi)}} \;\;\; (\gamma < \gamma^\star),
\end{equation}
consistently with the large-$\Omega$ expansion of Eq.~\eqref{eq:Delta_H_two_legs_exact_a}.
Once again we recognize the universal behavior if $\gamma \ll \gamma^\star$ since the second term at the denominator can be neglected, while we see a density dependence as $\gamma \to \gamma^\star$. 
In particular, the Hall imbalance at exactly the critical point is $\Delta_H(\gamma^\star) = -[1-\cos^2{\frac{n\pi}{2}}]^{-1}$ and it diverges as $-4/(\pi n)^2$ when $n\to 0$, consistently Eq.~\eqref{eq:deltaH_exact_2leg_small_n} at large $\Omega$.
If on the other hand $\gamma > \gamma^\star$, there are 4 Fermi points and the sums over $k$ are given in Eq.~\eqref{eq:integrals_4_Fermi_points}, with $\alpha-\beta = n$.
The Hall imbalance reads:
\begin{equation}
    \Delta_H = -\frac{\frac{\pi-\gamma}{\pi-\gamma^\star}}{1-\frac{(\pi-\gamma)^2}{(\pi-\gamma^\star)^2} \cos^2{\frac{n\pi}{2}}} \;\;\; (\gamma > \gamma^\star),
\end{equation}
which manifestly vanishes at $\gamma=\pi$, with a linear behavior.
We emphasize that, while the exact result in Eq.~\eqref{eq:Delta_H_two_legs_exact_b} is given implicitly in terms of $\alpha$ and $\beta$, this result is fully explicit, as it only depends on $t$, $\Omega$, $\gamma$ and $n$.
The evolution of $\Delta_H$ across the Lifshitz transition for several values of the density in the limit $\Omega \gg t$ is shown in the top panel of Fig.~\ref{fig:2-3-legs-vary-n}.
Results for  $1<n<2$ can be obtained via a particle-hole transformation ($n\to 2-n$) changing the sign of $\Delta_H$.

\subsection{3-leg ladder}

Let us now consider a 3-leg ladder where only the lowest band is partially filled ($0 < n < 1$).
Once again, the only magic flux is $\pi$.
%, since the solutions of eq. \ref{eq:magic_equation} are $\gamma_0 = 2\pi \frac{x}{M+1} = \frac{x}{2}\pi$ with $x=1,2$ and $x\neq 1,3$ (i.e. only $x=2$, which leads to $\gamma_0=\pi$).
Contrary to the 2-leg ladder, here the band develops the extra Fermi points at $\pm \pi$ ($\beta=1$), so that $\alpha-\beta = n - 1$, and from Eq.~\eqref{eq:critical_equation} we obtain
%where the characteristic function is \textcolor{blue}{$f_0(\gamma) = \frac{\sqrt{2}\cot^2{\frac{\gamma}{2}}}{3+5\cos{\gamma}} \approx - \frac{(\pi-\gamma)^2}{4\sqrt{2}}$ if $\gamma \approx \pi$},
\begin{equation}
    \gamma^\star = \pi - 2^{5/4} \sqrt{\frac{t}{\Omega}} \sin{\frac{n\pi}{2}}.
    \label{eq:gamma_star_3leg}
\end{equation}
\begin{figure}
    \centering
    \includegraphics[width=0.95\linewidth]{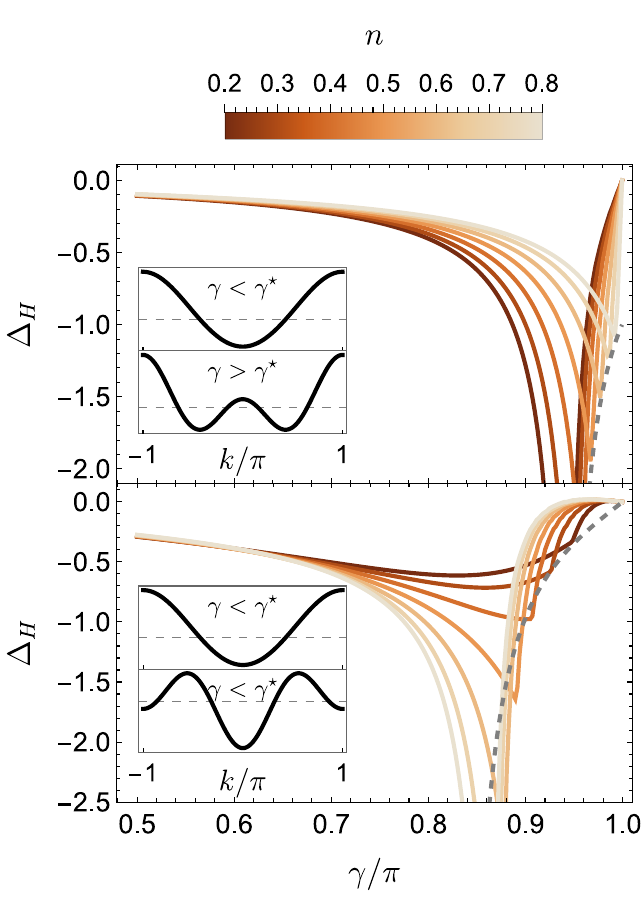}
    \caption{Hall imbalance for the 2-leg ladder (top panel) and 3-leg ladder (bottom panel) computed with exact numerical diagonalization for a large representative value of the synthetic tunneling $\Omega = 20t$.
    The displayed curves are obtained for different densities, with $N=100,\,150,\,200,\,250,\,300,\,350,\,400$ and $L=500$.
    We observe that $\Delta_H$ diverges at the Lifshitz point if $n\to0$ in the 2-leg ladder and if $n\to 1$ in the 3-leg ladder.
    The gray dashed lines are the set of points $(\gamma^\star, \Delta_H(\gamma^\star))$ as a function of the density predicted by perturbation theory.
    The insets show representative structures of the lowest band for $\gamma < \gamma^\star$ (2 Fermi points) and $\gamma > \gamma^\star$ (4 Fermi points).
    }
    \label{fig:2-3-legs-vary-n}
\end{figure}

\begin{figure*}[t]
    \centering
    \includegraphics[width=0.8\linewidth]{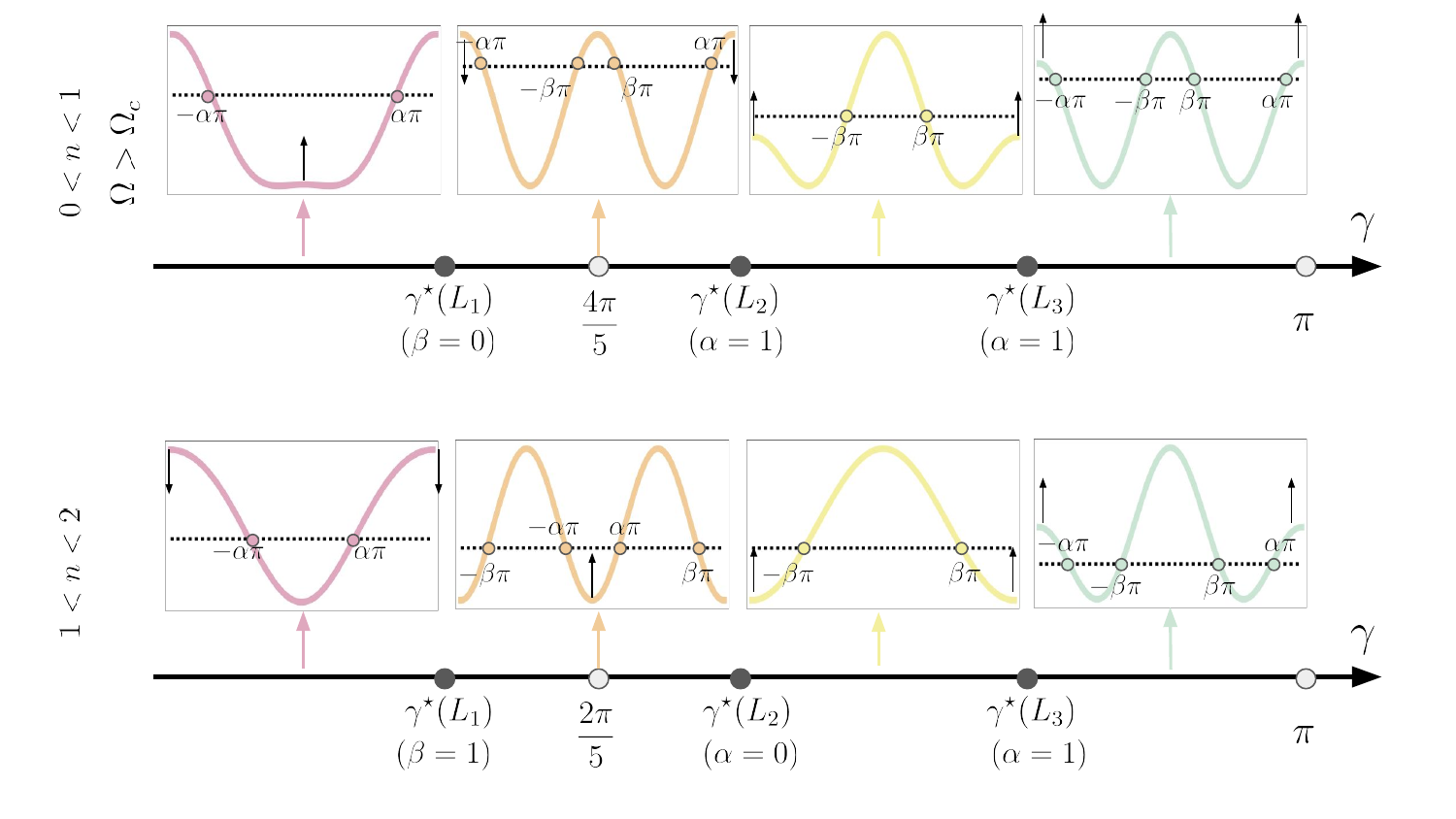}
    \caption{Evolution of the band structure of the 4-leg ladder as a function of $\gamma$ for a large value of $\Omega\gg t$. 
    The top panel shows the band $p=0$ in the case where 3 Lifshitz transitions occur ($\Omega > \Omega_c$). 
    If instead $\Omega < \Omega_c$ (not shown) the band edges at $k=\pm \pi$ never move below the Fermi energy, $L_2$ and $L_3$ are not present, and the phase with 4 Fermi points extends from $\gamma^\star(L_1)$ to $\pi$.
    The bottom panel shows the evolution of the band $p=1$.
    The figure also illustrates our convention on the labeling of the Fermi points, as discussed in Sec. \ref{sec:Perturbation_theory}. 
    }
    \label{fig:4-legs-bands}
\end{figure*}
We notice that the distance of $\gamma^*$ from $\pi$ is proportional to $\sqrt{t/\Omega}$, which is larger than the linear value ${t/\Omega}$ of the 2-leg ladder. Hence, the phase with 4-Fermi points is present in a wider region for the 3-leg ladder. Also the dependence on density is completely different. 
%Moreover, this interval becomes smaller as $n\to 0$, unlike in the 2-leg ladder, where it shrinks as $n\to 1$.
This is illustrated in Fig.~\ref{fig:2-3-legs-vary-n}, where we compare the exact Hall imbalance of the 2-leg and 3-leg ladders, and test the validity of Eq.~\eqref{eq:gamma_star_3leg} (see the dashed gray line).

If $\gamma < \gamma^\star$ there are 2 Fermi points and, combining Eqs.~\eqref{eq:Delta_H_perturbation_theory} and \eqref{eq:integrals_2_Fermi_points}, we get
\begin{equation}\label{eq:Delta_H_perturbative_3_legs_2_Fermi_points}
    \Delta_H = -\frac{\frac{4\sqrt{2}t}{\Omega}\tan{\frac{\gamma}{2}}}{1 - \frac{t}{\sqrt{2}\Omega} (3+5\cos{\gamma})\tan^2{\frac{\gamma}{2}} \cos{(n\pi)} } \;\;\; (\gamma < \gamma^\star).
\end{equation}
The structure of this equation is similar to what we have found for the 2-leg ladder in Eq.~\eqref{eq:Delta_H_perturbative_2_legs_2_Fermi_points}. 
When $\gamma \ll \gamma^\star$ we can neglect the second term in the denominator and we recover the universal behavior $\propto -\frac{t}{\Omega} \tan{\frac{\gamma}{2}}$ with a different prefactor ($4\sqrt{2}$ rather than $2$). Instead, for $\gamma \to \gamma^\star$ the Hall imbalance is density-dependent.
On the other hand, if $\gamma > \gamma^\star$ there are 4 Fermi points and we have to use Eq.~\eqref{eq:integrals_4_Fermi_points} for the sums over $k$, which leads to
\begin{equation}\label{eq:Delta_H_perturbative_3_legs_4_Fermi_points}
    \Delta_H = -\frac{2 \frac{(\pi-\gamma)^3}{(\pi-\gamma^\star)^2}}{1 - \frac{(\pi - \gamma)^4}{(\pi-\gamma^\star)^4} \sin^2{\frac{n\pi}{2}}} \;\;\; (\gamma > \gamma^\star).
\end{equation}
From Eq.~\eqref{eq:Delta_H_perturbative_3_legs_4_Fermi_points} we observe that $\Delta_H$ vanishes at $\gamma=\pi$ (as expected) with a cubic behavior, instead of the linear behavior observed in the 2-leg ladder. Furthermore, we can easily derive the Hall imbalance at the critical point: $\Delta_H(\gamma^\star) = -2^{9/4} \sqrt{\frac{t}{\Omega}} \frac{\sin{\frac{n\pi}{2}}}{1-\sin^2{\frac{n\pi}{2}}}$.
This value scales as $\sqrt{t/\Omega}$, hence for a generic density it is typically smaller than the 2-leg value, which instead does not depend on $t/\Omega$ at leading order.
The 3-leg result diverges as $\sim (n-1)^{-2}$ for  $n\to 1$, while the 2-leg ladder counterpart diverges for $n\to 0$.
Moreover, from the explicit expressions~\eqref{eq:Delta_H_perturbative_3_legs_2_Fermi_points} and~\eqref{eq:Delta_H_perturbative_3_legs_4_Fermi_points}, one can show that $\Delta_H$ behaves linearly close to the transition point $\gamma^\star$, with different slopes on the two sides, which diverge in the limit $n\to 1$. This feature is shared with the 2-leg case, except from the different regime in which the divergence occurs. 
However, unlike the 2-leg ladder,
%is the most sensitive to the critical point, because
where $\Delta_H(\gamma^\star)$ is the minimum value of the Hall imbalance, in the 3-leg ladder this is not necessarily the case.
For instance, looking at Fig.~\ref{fig:2-3-legs-vary-n} we observe that the critical Hall imbalance at low density is not the minimum value of the function.

If the lowest band is completely filled and the second band is only partially filled ($1<n<2$), then Eq.~\eqref{eq:Delta_H_perturbation_theory} predicts $\Delta_H \equiv 0$ for all fluxes, as discussed in Sec.~\ref{sec:Perturbation_theory}.
From the numerical calculation, we find that this result is correct for most fluxes, with the exception of a rather narrow interval of $\gamma$ where this approximation fails and the Hall imbalance has a non-trivial behavior. Deviations are a consequence of higher-order corrections in $t/\Omega$, which become relevant at specific combinations of density and fluxes, when lower-order terms are vanishingly small.
More consistent analytical results could thus be obtained by expanding the expression for the central band up to terms of order $(t/\Omega)^3$ and by correcting Eq.~\eqref{eq:Delta_H_perturbation_theory} accordingly.
%The reason is that, as a consequence of the band-reflection symmetry of the coefficients $C_{pp^\prime\,\sigma}$, the second-order term in the perturbative expansion of the bands vanishes identically and there are no terms contributing to the polarization, i.e. with finite mixed second derivative $\partial^2_{\phi\nu}E_0$.
%More reliable results can be obtained by keeping higher-order terms in the perturbative expansion of the bands.
Finally, if $2<n<3$ we can simply replace $n\to 3-n$ and change the global sign in Eqs.~\eqref{eq:Delta_H_perturbative_3_legs_2_Fermi_points} and \eqref{eq:Delta_H_perturbative_3_legs_4_Fermi_points}.

\subsection{4-leg ladder}
%As a last example, we discuss the Hall response of a 4-legs ladder, which is particularly interesting because on the one hand it discloses a rich physical scenario and on the other hand it proves the robustness of the perturbative formalism.
As a last example, we discuss the Hall response of a 4-leg ladder, which showcases several differences with respect to the $M=2$ or $3$ cases, allowing us to illustrate more aspects of the general formalism.
The first difference is already present by considering a partial occupation of the lowest band, i.e., $0<n<1$. Contrary to the 2- and 3-leg ladder, we encounter for the first time two magic fluxes: $\gamma_0=\frac{4\pi}{5}$ and $\gamma_0=\pi$.
%To begin with, let us consider the case $0<n<1$ where now, contrary to the 2-legs and 3-legs ladder, we find two magic fluxes: $\frac{4\pi}{5}$ and $\pi$.
%Contrary to the 3-legs and 2-legs ladder, here we have another magic flux where the Hall imbalance vanishes besides the symmetry-protected flux $\pi$.
The characteristic function of the band $f_0(\gamma)$ is positive in the interval $[0, \frac{4\pi}{5})$ and negative in $(\frac{4\pi}{5},\pi)$, it has a global minimum for $\gamma=\gamma_{\text{min}}$ (where $\frac{4\pi}{5}<\gamma_{\text{min}}<\pi$) and it is linear around the magic fluxes.
%, in particular $f_0(\gamma) \approx \frac{\sqrt{5-2\sqrt{5}}}{4} (\frac{4\pi}{5}-\gamma)$ if $\gamma \approx \frac{4\pi}{5}$ and $f_0(\gamma) \approx \frac{\sqrt{5}-3}{8\sqrt{5}} (\pi - \gamma)$ if $\gamma \approx \pi$.

The evolution of the band shape with respect to $\gamma$ for large values of $\Omega$ sketched in Fig.~\ref{fig:4-legs-bands} reveals the possible presence of up to 3 Lifshitz transitions, where $L_1$ is characterized by $\beta = 0$, while $L_2$ and $L_3$ by $\alpha = 1$.
Inserting these conditions in Eq.~\eqref{eq:critical_equation} and using the expansion of $f_0(\gamma)$ around the magic fluxes, we get the three critical fluxes:
\begin{subequations}
\label{eq:critical_fluxes_4_legs_p=0}
\begin{align}
    \gamma^\star(L_1) &= \frac{4\pi}{5} - \frac{4}{\sqrt{5-2\sqrt{5}}} \frac{t}{\Omega} \cos^2{\frac{n\pi}{2}},\\
    \gamma^\star(L_2) &= \frac{4\pi}{5} + \frac{4}{\sqrt{5-2\sqrt{5}}} \frac{t}{\Omega} \cos^2{\frac{n\pi}{2}},\\
    \gamma^\star(L_3) &= \pi - \frac{8\sqrt{5}}{3-\sqrt{5}} \frac{t}{\Omega} \cos^2{\frac{n\pi}{2}}.
\end{align}
\end{subequations}
The numerical solution reveals that $L_1$ exists for all the values of $\Omega$ such that the bands are gapped, while $L_2$ and $L_3$ are present only if $\Omega$ is sufficiently large, denoted by $\Omega>\Omega_c$ in Fig.~\ref{fig:4-legs-bands}.
{An estimate for $\Omega_c$ is obtained by solving the equation $\gamma^*(L_2)=\gamma^*(L_2)$, which gives $\Omega_c=t\omega[1+\cos(n\pi)]$, with a numerical prefactor $\omega\approx 23.02$. This estimate, can be translated into an expression for a critical density above which the transitions $L_2$ and $L_3$ are supported for a given $\Omega$: $n_c=\pi^-1\cos^{-1}(\Omega/t\omega-1)$.}
%This can be qualitatively explained in the framework of perturbation theory: the equation for $\gamma^\star(L_2)$ and $\gamma^\star(L_3)$, which is $f_0(\gamma) = - \frac{t}{\Omega} \cos^2{\frac{n\pi}{2}}$, has no solutions if the right-hand side is smaller than the global minimum of the characteristic function $f_0(\gamma_{\text{min}}) \approx -0.014$.
%Therefore, the system supports $L_2$ and $L_3$ only if $\Omega \gtrsim \frac{t}{|f_0(\gamma_{\text{min}})|} \cos^2{\frac{n\pi}{2}} \equiv \Omega_c$.
%On the other hand, the equation for $\gamma^\star(L_1)$, which is $f_0(\gamma) = + \frac{t}{\Omega} \cos^2{\frac{n\pi}{2}}$, always admits a solution because $f_0(\gamma)$ does not have an upper bound.

\begin{figure}[t]
    \centering
    \includegraphics[width=0.95\linewidth]{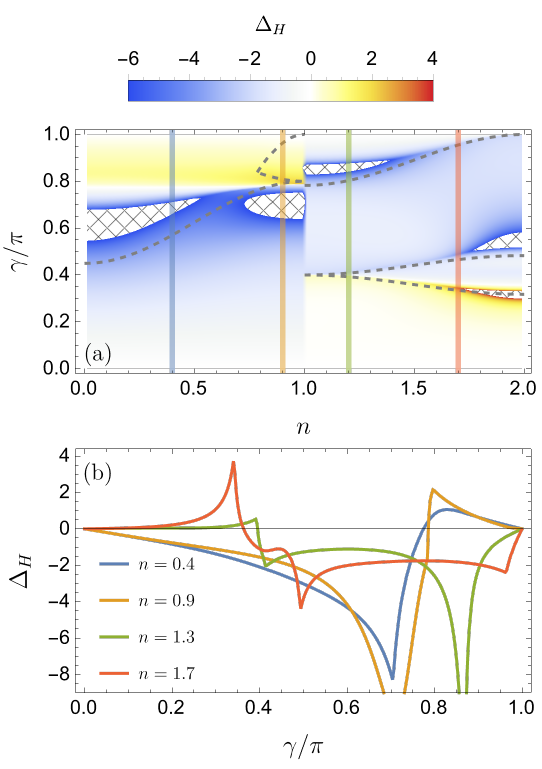}
    \caption{(a) Hall imbalance $\Delta_H$ of a 4-leg ladder computed numerically as a function of the particle density $n=N/L$ and the magnetic flux $\gamma/\pi$ for $\Omega = 5t$. 
    The Lifshitz transitions predicted with perturbation theory occur along the dashed lines. 
    If $1<n<2$ the three Lifshitz transitions, as well as their dependence on the density, are clearly visible from the sudden variation of $\Delta_H$.
    If $0<n<1$, since $\Omega_c$ depends on the density [see the discussion after Eq.~\eqref{eq:critical_fluxes_4_legs_p=0}], we have only $L_1$ for low densities where $\Omega < \Omega_c$ and all the three transitions for high densities where $\Omega > \Omega_c$. 
    The transitions $L_2$ and $L_3$ are poorly visible, because they are not characterized by a large variation of the Hall imbalance, cf.~(b).  
    The uncolored regions with a grid correspond to a large Hall imbalance exceeding the arbitrary maximum and minimum that we have set for the color scale. (b) Cuts at fixed densities, corresponding to the vertical lines in (a).
    }
    \label{fig:4Leg_n_gamma}
\end{figure}

The Hall imbalance can be computed explicitly from Eq.~\eqref{eq:Delta_H_perturbation_theory}; however the resulting expression is cumbersome, so here we only emphasize the main properties.
If $\gamma \ll \gamma^\star(L_1)$, we obtain a density-independent response of the form $\Delta_H \approx -\frac{t}{\Omega} \frac{a_1+a_2\cos{\gamma}}{a_3+a_4\cos{\gamma}} \tan{\frac{\gamma}{2}}$, where $a_i$ are positive constants. This is different from the behavior of the 2-leg and 3-leg ladders, where in this regime the only dependence on the flux is through $\tan{\frac{\gamma}{2}}$. 
If $\gamma \approx \frac{4\pi}{5}$, the Hall imbalance is linear in the flux: $\Delta_H \propto (\gamma-\frac{4\pi}{5})$, which means that $\Delta_H$ can take positive values after the first magic flux.
In fact, the numerical calculation shows that $\Delta_H>0$ not only close to $\gamma=\frac{4\pi}{5}$, but also in a large interval of fluxes and even for smaller values of $\Omega$, where predictions from perturbation theory are less accurate (see Fig.~\ref{fig:4Leg_n_gamma}). 
This opens the possibility to experimentally detect a change of sign in the Hall response of a 4-leg ladder upon increasing the flux at fixed density.

We can now move to the case $1<n<2$, where the band $p=1$ is partially filled.
The magic fluxes are different from those of the previous band and read $\gamma_0=\frac{2\pi}{5}$ and $\gamma_0=\pi$.
The characteristic function $f_1(\gamma)$ is neither lower-bounded nor upper-bounded, and is linear around the magic fluxes.
%: $f_1(\gamma) \approx \frac{\sqrt{5+2\sqrt{5}}}{4}(\gamma-\frac{2\pi}{5})$ if $\gamma \approx \frac{2\pi}{5}$ and $f_1(\gamma) \approx -\frac{\sqrt{5}+3}{8\sqrt{5}}(\pi-\gamma)$ if $\gamma\approx\pi$.
Looking at the evolution of the band upon increasing $\gamma$ in Fig.~\ref{fig:4-legs-bands}, we recognize three Lifshitz transitions, where $L_1$ is characterized by $\beta = 1$, $L_2$ by $\alpha = 0$ and $L_3$ by $\alpha = 1$.
Solving Eq.~\eqref{eq:critical_equation} with these conditions, we get
\begin{subequations}
\label{eq:critical_fluxes_4_legs_p=1}
\begin{align}
    \gamma^\star(L_1) &= \frac{2\pi}{5} - \frac{4}{\sqrt{5+2\sqrt{5}}} \frac{t}{\Omega} \sin^2{\frac{(n-1)\pi}{2}},\\
    \gamma^\star(L_2) &= \frac{2\pi}{5} + \frac{4}{\sqrt{5+2\sqrt{5}}} \frac{t}{\Omega} \sin^2{\frac{(n-1)\pi}{2}},\\
    \gamma^\star(L_3) &= \pi - \frac{8\sqrt{5}}{\sqrt{5}+3} \frac{t}{\Omega} \cos^2{\frac{(n-1)\pi}{2}}.
\end{align}
\end{subequations}
Since the function $f_1(\gamma)$ is not bounded, these solutions always exist, which qualitatively explains why in the numerical calculation we observe three transitions also for relatively small values of $\Omega$.
The Hall imbalance at small fluxes is qualitatively similar to the previous case: $\Delta_H \approx -\frac{t}{\Omega} \frac{b_1+b_2\cos{\gamma}}{b_3+b_4\cos{\gamma}} \tan{\frac{\gamma}{2}}$, where now the constants are different. In particular $b_1$ and $b_3$ are negative, while $b_2$ and $b_4$ are positive, resulting in $\Delta_H > 0$.
At the magic fluxes, the Hall imbalance is linear: $\Delta_H \propto (\frac{2\pi}{5}-\gamma)$ if $\gamma \approx \frac{2\pi}{5}$ and $\Delta_H \propto (\gamma-\pi)$ if $\gamma \approx \pi$. 
This means that once again the imbalance changes sign upon increasing the flux.
At the critical fluxes, $\Delta_H$ is non-differentiable, however it is not necessarily a maximum or a minimum (this depends on the specific Lifshitz point), similarly to the case of the 3-legs ladder.
%Nevertheless, in the intervals of $\gamma$ between a critical flux and a magic flux we observe a sudden variation of $\Delta_H$ from a finite and generally large value to $0$; such a rapid change helps to identify the critical fluxes in the plots.
All these results are summarized in Fig. \ref{fig:4Leg_n_gamma}, where we perform a numerical calculation of $\Delta_H$ at an intermediate value of $\Omega = 5t$.
By means of the perturbative approach discussed here we are able to qualitatively explain all the main features of the result: the sudden variation of $\Delta_H$ around the magic fluxes (where $\Delta_H=0$), the change of sign observed fixing $n$ and incrementing $\gamma$, or fixing a small $\gamma$ and varying $n$, and the evolution of the critical fluxes with the density [see Eqs.~\eqref{eq:critical_fluxes_4_legs_p=0}, \eqref{eq:critical_fluxes_4_legs_p=1}].

\begin{comment}
\begin{figure}
    \centering
    \includegraphics[width=\linewidth]{4leg_1vs3transitions.pdf}
    \caption{Caption}
    \label{fig:enter-label}
\end{figure}
\end{comment}

\section{Bond currents and vortex phase}
\label{sec:Bond_and_rung_currents}

It has been previously suggested that the Hall imbalance is sensitive to the so-called Meissner-vortex transition, which should be accompanied by divergencies in $\Delta_H$~\cite{Giamarchi_prl_2021}. Based on our analysis in Sec.~\ref{sec:Perturbation_theory}, we know that such divergencies only arise in limiting density regimes of almost empty or almost filled bands, one of which clearly mimics the bosonic case. The scenario is less evident in the fermionic case, which motivates us to investigate the site-resolved currents in detail
to get a complete picture of the behavior of the system.

On a bosonic 2-legs ladder, the flux-driven change of shape of the energy bands (analogue of a Lifshitz transition for fermions) leads to a significant change of the spatial pattern of persistent currents~\cite{Atala,Piraud2015Apr,DiDio2015Apr, Orignac_2016,Impertro2025Jun}. 
The Meissner phase occurs when the band has a single minimum and is characterized by uniform currents flowing along the legs only, and no current along the rungs.
Conversely, a vortex phase occurs when the band has a double minimum and it is characterized by significant spatial modulations of the currents, leading to a pattern of vortices alternated to antivortices or density-depleted regions.

Let us now consider an $M$-leg fermionic ladder with a finite number of lattice sites $L$, real PBC and gapped bands. 
Since at equilibrium the currents satisfy Kirchhoff's current law at every node, we can limit our discussion to the bond currents flowing along the real direction, which uniquely determine the value of the bond currents flowing along the rungs.
The bond current connecting the nodes $(j,\sigma)$ and $(j+1,\sigma)$ is defined as
\begin{align}\label{eq:bond_current_general}
    J_x(j,\sigma) =& -it e^{i\gamma\sigma} \langle c^{\dag}_{j\,\sigma} c_{j+1\,\sigma} \rangle + \mathrm{h.c.} \nonumber \\
    =& -\frac{it e^{i\sigma\gamma}}{L} \sum_{kk^\prime} \langle c^{\dag}_{k\sigma} c_{k^\prime\sigma} \rangle e^{-ikj} e^{ik^\prime(j+1)} + \mathrm{h.c.},
\end{align}
where we have Fourier-transformed the operators to momentum space.
%If for concreteness we assume that the corresponding system in the thermodynamic limit has two Fermi points and that $L$ is even (specular arguments hold if $L$ is odd), then $k=0$ is a sampled momentum state and because every band is symmetric under $k\to -k$, the many-body ground state is degenerate if $N$ is even and non-degenerate if $N$ is odd.
The expectation value $\langle c^{\dagger}_{k\sigma} c_{k^\prime \sigma} \rangle$ depends on the degeneracy of the many-body ground state.
In the absence of degeneracy, when two fermions populate the two degenerate quasi-momentum states in the $p$-th band $|\pm k_F\rangle \otimes |p\rangle$, we have $\langle c^{\dagger}_{k\sigma} c_{k^\prime \sigma} \rangle = \delta_{kk^\prime} n_{k\sigma}$ if $|k| \leq k_F$, where $n_{k\sigma}$ is the number of particles in the quasi-momentum state $k$ and leg $\sigma$, and $0$ otherwise. 
The resulting leg current is spatially uniform, and there is no current flowing along the rungs (Meissner phase).

On the other hand, in the presence of a two-fold degeneracy, the ground state can be constructed as a generic quantum superposition $a|+\rangle + b|-\rangle$, where $a$ and $b$ are complex numbers and $|\pm \rangle$ represent degenerate many-body states where one fermion at the Fermi level populates the single-particle states $|\pm k_F\rangle \otimes |p\rangle$, respectively.
The equal-weight linear combination with $a = b = 1/\sqrt{2}$ is particularly relevant for experimental implementations (see~\cite{Atala} for the bosonic case), as it mimics the ground state of the corresponding system with open boundary conditions and a weak harmonic trap in the bulk region of the ladder.
However, here we keep the discussion more general with arbitrary $a$ and $b$.
With this ground state, there is an additional contribution to the expectation value in Eq.~\eqref{eq:bond_current_general}. By writing $\langle  c^{\dagger}_{k_F\sigma} c_{-k_F \sigma}\rangle = A e^{i\theta} \propto a^\star b$ and $\theta \in [0,2\pi]$, the resulting site-resolved bond current has a uniform contribution and an oscillating part:
\begin{align}\label{eq:bond_current_L_even_2_Fermi_points}
    J_x(j,\sigma) &= \frac{2t}{L} \sum_{|k| \leq k_F} w_k \sin{(k+\sigma\gamma)} n_{k\sigma} \nonumber\\
    &+ \frac{4t}{L} A \sin{(\sigma\gamma)} \cos{(2k_Fj + k_F - \theta)},
\end{align}
where $w_k = 1$ if $|k| < k_F$, $w_{k_F} = |a|^2$ and $w_{-k_F} = |b|^2$. 
The states $|\pm k_F\rangle \otimes |p\rangle$ are defined up to a phase factor, which can be absorbed into $\theta$, the relative phase between $a$ and $b$, which is effectively a gauge degree of freedom.
%This means that $\theta$ is a gauge degree of freedom which is constrained to be an integer multiple of $2k_F$ .
The wavelength of the spatial modulation is $\lambda = \pi/k_F$; however, since the function $J_x(j,\sigma)$ is defined on a discrete domain $j=1,\dots,L$, the observed period is indeed $\lambda$ only if the latter is an integer, otherwise it is different due to the stroboscopic effect. 
We emphasize that Eq.~\eqref{eq:bond_current_L_even_2_Fermi_points} is fully consistent with the requirement of real PBC no matter the shape of the band (as long as $\gamma$ is an integer multiple of $4\pi/L$), because $2k_F L$ is guaranteed to be an integer multiple of $2\pi$, and consequently the number of repeating periods $L/\lambda$ is always an integer.
Moreover, the amplitude of the modulation is $\propto 1/L$, while the uniform contribution coming from the Fermi sea is $\mathcal{O}(L^0)$ for a generic density. 
Therefore, on a relatively large system, the spatial modulation is typically much smaller than the uniform background current and the formation of vortex-antivortex pairs is inhibited. 
If instead the density of particles or holes is very low, the background current is comparable to the modulation, if not smaller, and the current may change sign along the legs, which could lead to vortex-antivortex patterns.

\begin{figure*}
    \centering
    \includegraphics[width=0.7\linewidth]{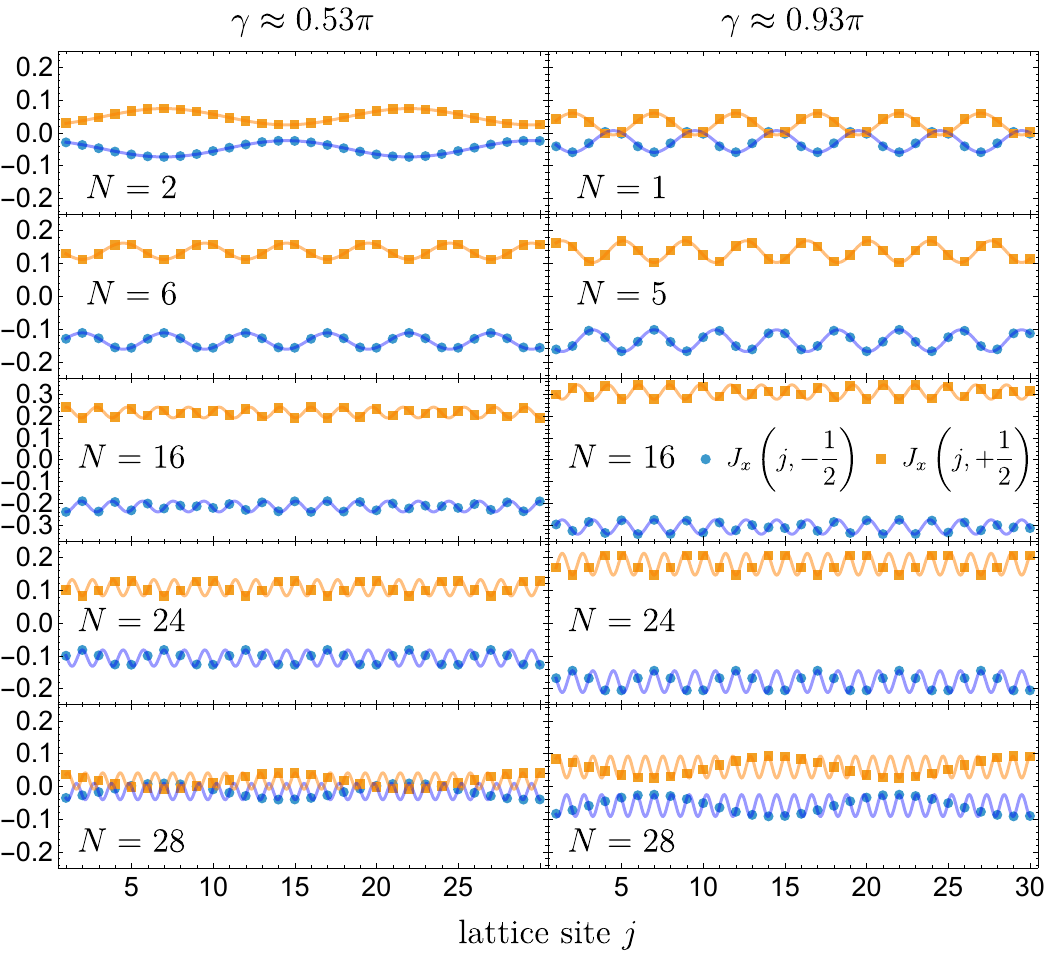}
    \caption{Evolution of the pattern of leg currents in a 2-legs ladder with $L=30$ sites as a function of the particle number $N$ for a band with a single minimum (left panel) and with a double minimum (right panel), computed from Eq.~\eqref{eq:bond_current_L_even_2_Fermi_points} using $a=b=\frac{1}{\sqrt{2}}$ and $\Omega = 15t$. 
    The solid lines are guides for the eye obtained by promoting $j$ to a real number in Eq.~\eqref{eq:bond_current_L_even_2_Fermi_points}. 
    The chosen fluxes are commensurate with PBC, i.e. they are in the form $\gamma = \frac{4\pi m}{L}$ with $m=4$ (left) and $m=7$ (right). }
    \label{fig:vortex_Meissner}
\end{figure*}
A system with finite $L$ and generic $\Omega$ and $\gamma$ has only two Fermi points regardless of the shape of the bands, and the corresponding ground state is either non-degenerate or two-fold degenerate depending on the parity of $N$ and $L$.
Nevertheless, we can still classify the Lifshitz phases by the number of intersections between the Fermi energy of the finite-size system and the corresponding band in the thermodynamic limit.
In the phase with 4 intersections $\pm k_1$ and $\pm k_2$ ($k_1 < k_2$), only two of them correspond to allowed quasi-momentum states in the finite-size system and those are the Fermi momenta. 
Upon evolving one of the parameters (e.g. $\gamma$), the Fermi momenta can suddenly change from $k_1$ to $k_2$ or vice versa, leading to an unpredictable wavelength in this phase.
On the other hand, in the phase with two intersections and in presence of a degenerate ground state, the Fermi momentum is either $k_F = (n-p)\pi$ for a band with the minimum in $k=0$ or $k_F = (1-n+p)\pi$ for a band with minimum in $k=-\pi$. 
The number of repeating periods of the current modulation equals the number of fermions in the $p$-th band $N-pL$ in the former case and the number of holes $L - (N-pL)$ in the latter.
%On a system with finite $L$, due to the $k\to -k$ symmetry of the bands, the many-body ground state is either non-degenerate or two-fold degenerate regardless of the shape of the band.
%Nevertheless, we can still classify the Lifshitz phases by the number of intersections between the Fermi energy of the finite-size system and the corresponding band in the thermodynamic limit.
%\sout{In the phase with 2 Fermi points (in this broader sense), we have $k_F = n\pi$, $\lambda = 1/n$, and the observed number of oscillations coincides with the particle number $N$.
%Instead,} In the phase with 4 Fermi points (in this broader sense), the quasi-momentum states $\alpha\pi$ and $\beta \pi$ defined in the thermodynamic limit are no longer degenerate and the true Fermi momentum $k_F$ descends from one of the two (which one depends on the density and on the specific Hamiltonian parameters).

As an example, in Fig.~\ref{fig:vortex_Meissner} we show the pattern of persistent currents in the ground state of a 2-leg ladder with $L=30$ sites for different values of the particle number $N$, in the limit of large $\Omega$.
In particular, we consider two representative values of the flux commensurate with PBC: $\gamma \approx 0.53\pi$ for a band with a single minimum and $\gamma=0.93\pi$ for a band with a double minimum. 
The presented values of $N$ span the whole range of densities and they all have a parity that ensures a two-fold degeneracy of the ground state manifold, from which we select the state with $a = b = 1/\sqrt{2}$.
In the figure, we observe several important features of Eq.~\eqref{eq:bond_current_L_even_2_Fermi_points}.
First of all, in a band with a single minimum, when the number of particles is such that $\frac{L}{N} \equiv \lambda \in \mathbb{N}$ (e.g., $N=2,\,6$), we observe a commensurate oscillation with exactly $N$ repeated periods. 
If $N$ does not divide $L$ exactly but $L-N$ does (e.g., $N=24,\,28$), even though $\lambda$ is not an integer multiple of the lattice spacing, the effectively observed period due to the stroboscopic effect is $\frac{L}{L-N}$, which can be interpreted as a modulation provided by the holes. 
In the opposite case of a band with double minimum (and a local maximum in $k=0$), if the Fermi level is higher than $\varepsilon_p(k=0)$, then $k_F = n\pi$ as in the previous case, otherwise it is non-trivially related to the density and the resulting periodicity is not obvious. 
This behavior is reflected in the right-panel of the figure, where for $N=16,\,24,\,28$ the Fermi level is above $\varepsilon_p(k=0)$ and the observed modulation is comparable to the previous case; while for instance with $N=5$ we do not see $5$ repeated periods ---as expected for a smaller $\gamma$--- and instead we observe $8$ periods.

Finally, at intermediate density, the uniform contribution to the current coming from the Fermi sea is large and the modulation $\propto 1/L$ mildly perturbs it, essentially resulting in an imperfect Meissner phase; whereas in the limit of small or large densities the modulation is comparable to the background current and the ground state displays a vortex phase. 
Depending on the specific parameters, the vortex phase can consist of vortex-antivortex pairs if the leg currents periodically change sign across the leg; or simply by isolated vortices alternated to depleted regions if the currents periodically vanish across the leg.
In the specific case of a single fermion, $N=1$, we recover the phenomenology of the bosonic 2-leg ladder \cite{Atala,Fermionic_ladder_with_qbits}, with a flux-driven Meissner-vortex transition at $\gamma = \gamma_c$. 

In conclusion, as for the bosonic case, the mechanism providing a current modulation in a fermionic ladder is the ground-state degeneracy and in particular the possibility to form a quantum superposition of the quasi-momentum states $|\pm k_F\rangle \otimes |p\rangle$ with non-vanishing coefficients.
However, in contrast to the bosonic case, the degeneracy is related not only to the Lifshitz phase but also to the relative parity of the number of sites and particles.
So, for a density-driven Lifshitz transition at fixed $\gamma$, there is no connection between ground-state degeneracy and Lifshitz phase. 
Moreover, the Pauli principle affects the particle distribution in the quasi-momentum states, providing a relatively large background current and effectively hiding the modulation.
Hence if the Lifshitz transition is flux-driven with a fixed density, the ground-state degeneracy changes at the critical point, but the background uniform current is not much affected, and there is no clear transition from a Meissner to a vortex phase.

\section{Conclusions and Outlook}
\label{sec:Conclusions}

In this work we have studied the Hall effect and the pattern of persistent equilibrium currents in fermionic ladders pierced by a synthetic gauge flux that can be realized in cold-atom platforms.

Previous influential work has focused on the low-flux regime, where the Hall imbalance $\Delta_H$ has been shown to be independent of the particle density \cite{Filippone_Giamarchi_Hall_imbalance, Giamarchi_prl_2021, Fallani_universal_Hall_response}, a behavior which is also robust with respect to the introduction of a $SU(M)$-symmetric Hubbard-like interaction, thus defining a universal behavior. A density dependence was indeed  found in some particular cases~\cite{Giamarchi_prl_2021}, but a comprehensive point of view on the role of density was still lacking.

Here we have filled this gap through an extensive analytical study of the non-interacting system that shows significant deviations from the universal behavior at large values of the flux. 
%More specifically, applying a second-order perturbative expansion for large values of the Raman coupling $\Omega$, we have motivated 

Notably, we have identified  ``magic fluxes", at which the Hall imbalance vanishes, and critical Lifshitz fluxes, where the number of Fermi points changes and consequently $\Delta_H$ is not analytical and it displays divergences.
The specific values of the Lifshitz points significantly depend on the density, and so does the Hall imbalance in the neighborhood of these points. 
Interestingly, exactly at the critical points the divergence of $\Delta_H$ can occur either in the zero-density limit or in the opposite large density limit, depending on the specific character of the energy band.

These results can be rationalized in simple  perturbative expressions valid in the large $\Omega$ regime. Comparison with numerical exact solutions shows that these expressions hold down to intermediate values of $\Omega$.
Therefore, this approach provides a relatively simple analytical framework to reliably investigate the Hall effect in a wide range of parameters, identifying a variety of non-trivial phenomena.

Our results provide a complete map of the deviations with respect to the universal regime, for case studies with $M=2,\,3,\,4$ flavors, which are particularly interesting because such small ladders lie well within the reach of modern experiments with ultracold alkaline-earth-like fermionic atoms.

Finally we have complemented this analysis by investigating the potential connections of the Hall response across a Lifshitz transition with the variation of the spatial pattern of ground-state currents. 
In particular we have found that, as opposed to the bosonic case, in fermionic ladders there is no strict connection between a Lifshitz transition and a Meissner-vortex phase transition. 
The first reason is that a spatial modulation of the leg current is related to the ground-state degeneracy, which, in a finite system depends on the parity of the particle number and not on the specific Lifshitz phase.
The second reason is that the Fermi sea provides a generally large background current that effectively hides the modulation, making it more difficult to define a proper distinction between Meissner and vortex phases.

The solid understanding of the Hall response beyond the universal regime in a non-interacting system is a fundamental building block for further investigations, in particular concerning the role of interactions in the many features of the non-universal response. 
Recent progress in experimental techniques to achieve an accurate control on the number of particles loaded into an optical lattice of small size offer the possibility to investigate the role of density in the Hall imbalance and pave the way to the detection of spatial current patterns.
Exciting perspectives are also offered by the experimental possibility to realize multiorbital ladders, for example by exploiting the long-lived electronic states of ultracold fermionic alkaline-earth-like atoms~\cite{Gorshkov_SU(N),Scazza_spin_exchange,Amaricci_PRA_2025}, where the interaction includes inter-orbital couplings and it is expected to disclose different scenarios.

A further interesting direction would be to investigate the Hall response in the presence of interplay between gauge flux and geometrical frustration~\cite{Beradze2023Jan,Beradze2023Aug,Li2023Nov,Halati2025Feb,Halati2025May}.

\acknowledgments

We acknowledge Stefania De Palo
%, Francesco Scazza and Roberta Citro
for fruitful discussions and financial support from the National Recovery
and Resilience Plan PNRR MUR Project No.\,CN00000013-ICSC and MUR Project No.\,PE0000023-NQSTI, by MUR via PRIN
2020 (Prot.\,2020JLZ52N-002) and PRIN 2022 (Prot.\,20228YCYY7) programs. 

\begin{appendix}
\section{Hall coefficient}
\label{sec:Hall_coefficient}

In the classical theory of the Hall effect, a two-dimensional rectangular slab of a conducting material with sizes $L_x$ and $L_y$ is pierced by a perpendicular magnetic field $B$ and is connected to a current generator, which induces a current density $j_x$ (that we assume conventionally parallel to the size of length $L_x$).
At equilibrium, the system develops an electric field $E_y$ parallel to the size with length $L_y$ and the Hall response is conventionally characterized by the Hall coefficient $R_H = \frac{E_y}{j_xB} = \frac{1}{n_c q}$, where $q$ is the charge of the carriers and $n_c$ is their number per unit of area.

In  the physical dimension we take PBC and $L_x = La$, where $a$ is the lattice spacing and $L$ the number of sites. We can formally introduce an arbitrary value $b$ for the lattice spacing along the synthetic direction and write $L_y = (M-1)b$, where $M-1$ is the number of synthetic bonds per lattice site.
Furthermore, in a synthetic ladder the particles are electrically neutral, but in order to make a correspondence to the classical case, we can introduce a fictitious charge $q$.
With these prescriptions, the quantum analogue of the current density is $j_x \to \frac{q}{(M-1)b} \frac{J_x}{hL}$, where in the linear response regime $J_x \approx \phi \cdot \partial^2_{\phi\phi} E_0|_{\phi=\nu=0}$, $h$ is the Planck's constant and $\frac{J_x}{hL}$ represents the average particle current per bond in the real direction.
The analogue of the magnetic field is $B \to \frac{h}{q}\frac{\gamma}{ab}$, where $\frac{h}{q}$ is the flux quantum, $\gamma$ is the number of quanta per plaquette (which is not limited to integer values in this case) and $ab$ is the area of a plaquette.
Finally, the analogue of $E_y$ is proportional to the value of the external field $\nu$ that is necessary to compensate the Hall-induced polarization~\cite{Zhou2025Nov}, i.e., $\nu_0$ such that $P_y(\phi, \nu_0) = 0$. 
In the linear response regime, the polarization is $P_y(\phi,\nu) \approx \phi\cdot \partial^2_{\phi\nu}E_0|_{\phi=\nu=0} + \nu\cdot \partial^2_{\nu\nu}E_0|_{\phi=\nu=0}$, which yields
\begin{equation}
    \nu_0 = - \frac{\partial^2_{\phi\nu}E_0|_{\phi=\nu=0}}{\partial^2_{\nu\nu}E_0|_{\phi=\nu=0}} \phi.
\end{equation}
More precisely, since $2\nu_0(M-1)$ represents the potential energy difference across the synthetic direction that makes the polarization vanish, the corresponding electric field is obtained dividing by $q$ and by $(M-1)b$, hence: $E_y \to \frac{2\nu_0}{bq}$.  
Using these prescriptions in the classical definition of $R_H$ we obtain $R_H = \frac{E_y}{j_xB} \to \frac{L(M-1)ab}{q} \frac{2\nu_0}{\gamma J_x}$.
However, since $b$ and $q$ are not well defined on a ladder, we can simply remove the factor $\frac{b}{q}$ in the definition of the Hall coefficient for a synthetic ladder, which is:
\begin{align}\label{eq:RH_definition}
    R_H = & - 2L(M-1)a \frac{\partial^2_{\phi\nu}E_0|_{\phi=\nu=0}}{\gamma \partial^2_{\phi\phi} E_0|_{\phi=\nu=0} \cdot \partial^2_{\nu\nu} E_0|_{\phi=\nu=0}} \nonumber\\
    = & - \frac{L(M-1)a \Delta_H}{t\gamma\partial^2_{\nu\nu} E_0|_{\phi=\nu=0}},
\end{align}
where we have used Eq.~\eqref{eq:Delta_H_definition}.

The intrinsic polarizability $\partial^2_{\nu\nu} E_0|_{\phi=\nu=0}$ can be computed within the perturbative expansion for $\Omega \gg t$ discussed in Sec.~\ref{sec:Perturbation_theory}.
Assuming that $p<n<p+1$, (i.e., that the $p$-th band is partially populated while all the bands from $0$ to $p-1$ are completely filled), we get 
\begin{align}\label{eq:intrinsic_polarizability_perturbation_theory}
    \partial^2_{\nu\nu} E_0|_{\phi=\nu=0} &= \sum_{p^\prime \neq p} \frac{8(N-pL)}{\varepsilon_p^{(0)} - \varepsilon_{p^\prime}^{(0)}} \left( \sum_{\sigma} \sigma C_{pp^\prime \sigma} \right)^2 \nonumber\\ &+ \sum_{p^{\prime\prime} = 0}^{p-1} \sum_{p^\prime \neq p^{\prime\prime}} \frac{8L}{\varepsilon_{p^{\prime\prime}}^{(0)} - \varepsilon_{p^\prime}^{(0)}} \left( \sum_{\sigma} \sigma C_{p^{\prime\prime}p^\prime \sigma} \right)^2. 
\end{align}
Since this does not depend on $\gamma$, the Hall coefficient $R_H$ qualitatively behaves as the Hall imbalance $\Delta_H$: it vanishes at the magic fluxes and changes slope at the Lifshitz points.
Furthermore, it is interesting to observe that in the limit $\gamma \to 0$ and considering the particular case $0<n<1$, a fermionic ladder has a \emph{classical} Hall response~\cite{Filippone_Giamarchi_Hall_imbalance}. 
To see this, we first expand Eq.~\eqref{eq:Delta_H_perturbation_theory} at small fluxes, observing that the second term at the denominator can be neglected. Then we use Eqs.~\eqref{eq:RH_definition}, \eqref{eq:intrinsic_polarizability_perturbation_theory}, and we get $R_H = \frac{L(M-1)a}{N}$.
To see how this compares to the classical result, we can reintroduce the factor $\frac{b}{q}$ and recognize that the carrier density is $n_c = \frac{N}{L(M-1)ab}$, which immediately yields $R_H \to \frac{1}{n_c q}$.

\end{appendix}

\bibliography{refs}

\end{document}